\documentclass[twocolumn,trackchanges]{aastex7}
\usepackage{algorithm2e}
\usepackage{orcidlink}
\usepackage{amssymb}
\usepackage{newtxtext,newtxmath}
\usepackage{siunitx}
\usepackage[utf8]{inputenc}
\usepackage[T1]{fontenc}
\usepackage{CJKutf8}
\usepackage{booktabs}

\usepackage{graphicx}	
\usepackage{amsmath}	

\SetKwComment{Comment}{/* }{ */}

\begin{document}

\title{The Universe is Odd}
\author[gname=Shen S.]{Shiyin Shen~\begin{CJK}{UTF8}{gbsn}(沈世银)\end{CJK}
~\orcidlink{0000-0002-3073-5871},\thanks{Email: \href{mailto:ssy@shao.ac.cn}{ssy@shao.ac.cn}}}
\affiliation{Shanghai Astronomical Observatory,  Chinese Academy of Sciences, Shanghai, 200030, China}
\email{ssy@shao.ac.cn},\thanks{Email: \href{mailto:ssy@shao.ac.cn}{ssy@shao.ac.cn}}

\author[gname=Li N.]{Nan Li~\begin{CJK}{UTF8}{gbsn}(李楠)\end{CJK}
~\orcidlink{0000-0001-6800-7389},\thanks{Email: \href{nan.li@nao.cas.cn}{nan.li@nao.cas.cn}}}
\affiliation{National Astronomical Observatories, Chinese Academy of Sciences, Beijing, 100101 China}
\email{nan.li@nao.cas.cn},\thanks{Email: \href{nan.li@nao.cas.cn}{nan.li@nao.cas.cn}}

\begin{abstract}


The cosmological principle posits that the universe does not exhibit any specific preference for position or direction. However, it remains unclear whether the universe has a distinct preference for parity—whether certain properties are more likely to be classified as even or odd. In this study, we analyze the largest available galaxy group catalogs to explore this hypothesis: specifically, whether the number of galaxies within a galaxy group or cluster is more likely to be odd or even. Our findings convincingly indicate that the universe indeed favors odd numbers, with results achieving a significance level well above the $4.1-\sigma$ threshold.

\end{abstract}

\keywords{}

\section{Introduction}
The cosmological principle, often referred to as the Copernican Principle, is one of the foundational pillars of modern cosmology. It posits that the universe is homogeneous and isotropic on large scales, meaning that there is no preferred location or direction in the universe. This principle has been instrumental in shaping our understanding of the universe's large-scale structure and its evolution over time. By assuming homogeneity and isotropy, astronomers have been able to develop robust theoretical models, such as the Friedmann-Lema\^itre-Robertson-Walker (FLRW) metric, which describes the expansion of the universe. Observational data, such as those from large-scale structure surveys and cosmic microwave background missions, have also shown strong support for the homogeneity and isotropy of the universe\footnote{This paragraph is mainly written by AI.}. 

Nevertheless, the universe does not consistently exhibit randomness and symmetry throughout. For instance, several studies have provided evidence indicating that the distribution of spiral galaxies' spin directions lacks both randomness and symmetry. A greater proportion of galaxies in the northern hemisphere rotate counterclockwise, whereas the southern hemisphere exhibits the reverse trend\citep[][but also see \citet{Patel2024}]{Shamir22}. Besides the chirality shown by the spiral arms, it is not clear whether the universe has a preference for other binary properties, e.g. \textit{positive or negative, odd or even}, i.e. \textit{to be or not to be}.

According to the current standard cold dark matter cosmology scenario, galaxies are formed inside dark matter halos and then grow through the hierarchical merging process. Because the merging process is non-linear and the merging time-scale is quite long,  a dark matter halo typically hosts more than one galaxy. Therefore,  for a random universe with no preferences, the number of member galaxies inside a  galaxy group (dark matter halo) should also be random, with equal probability of being odd or even. Whether this hypothesis is true? Is it possible that our universe has a preference for parity? This paper aims to provide a quick answer to this nontrivial question. 

The structure of this paper is as follows. Sect. \ref{section:data} introduces the data used in this work. In Sect. \ref{section:method}, we detail our method to calculate the parity of our universe using the above datasets, and present our findings in Sect. \ref{section:result}. Lastly, our conclusions are drawn in Sect. \ref{section:conclusion}. 


\section{Data}
\label{section:data}
We utilize the group catalog derived from the DESI Legacy Imaging Surveys DR9 by \citet{Yang2021}. This catalog represents the most extensive galaxy group catalog currently available. It encompasses two sky regions: the south galactic cap (SGC) and the north galactic cap (NGC). For each of these regions, we select groups that comprise at least three members \footnote{\begin{CJK}{UTF8}{gbsn}一生二，二生三，三生万物。\end{CJK}}, yielding a comprehensive sample size of over 5.7 million groups and over 30 million group members. \textit{Such extensive numbers permit a statistically robust analysis.}

\section{method}
\label{section:method}
We first count the number of group members for each group. Then, we sort the groups into two categories: the groups with odd members and the groups with even members.  Finally, we calculate the total number of groups with odd members, denoted as $N_{odd}$, and those with even members, denoted as $N_{even}$.The  algorithm flow and pseudo-code is outlined in Algorithm 1.

\RestyleAlgo{ruled}
\begin{algorithm}
\caption{\small{Algorithm of counting groups with odd and even members.}}
\SetAlgoLined
\KwIn{Member ID of DESI DR9 group members}
\KwOut{$N_{odd}$, $N_{even}$}
\Comment{count the number of group members $N_{mem}$ for each group}
$N_{odd} = 0$\;
$N_{even} = 0$\;
\For{$Group \in All\,Groups$} {
    \eIf{$N_{mem}\,\%\,2\,\neq\,0$} {
        $N_{odd}$ += 1\;
    }
    {
        $N_{even}$ += 1\; 
    }
}
\end{algorithm}

\begin{figure}[htb!]
\centering
\includegraphics[width=\columnwidth]{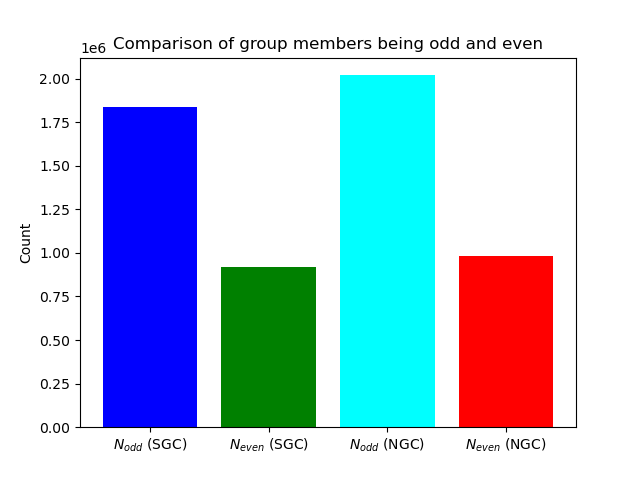}
    \caption{The numbers of DESI DR9 groups with odd and even group members in the the south galactic cap(SGC) and the north galactic cap(NGC) regions respectively.}
    \label{fig: odd_even}
\end{figure}

\section{Result}
\label{section:result}

For DESI DR9 groups with three or more members, there are 1,839,384 odd-member groups in the SGC area and 2,019,782  in the NGC area. In contrast, the counts for even-member groups are only  918,633  and 983,507 respectively. Figure 1 presents these findings with a bar plot. Clearly, in both regions, the numbers of odd-member groups  are more than twice the even-member groups, highlighting a significant discrepancy. With simple Poisson statistics, the significance of this discrepancy is much higher than 4.1-sigma level.

\section{Conclusion}
\label{section:conclusion}
There is only \textbf{One} universe, and \textbf{this Universe is Odd.}\\

\section*{Data Availability}
\label{sec: public}
 The code of this study is available at \url{http://cluster.shao.ac.cn/~shen/odd_even.ipynb} and the DESI DR9 group catalog is available at \url{https://gax.sjtu.edu.cn/data/data1/DESI_DR9/DESIDR9_member.tar.gz}.

\begin{acknowledgments}
This study is not supported by any scientific grants. SS thanks Dr. Lu Li for the not very parallel but maybe competitive project. NL made no academic contribution to this study and is included here solely to ensure the number of authors to be \textbf{even}. 
\end{acknowledgments}

\bibliography{sample7}{}
\bibliographystyle{aasjournalv7}

\end{document}